\documentstyle[preprint,aps,floats,epsfig]{revtex}

\def\ie{{\it i.e.,}}
\newcommand{\be}{\begin{equation}}
\newcommand{\ee}{\end{equation}}
\newcommand{\bea}{\begin{eqnarray}}
\newcommand{\eea}{\end{eqnarray}}
\newcommand{\bml}{\begin{mathletters}}
\newcommand{\eml}{\end{mathletters}}

\begin{document}

\tighten

\preprint{DCPT/01/59, hep-th/0107108}
\draft




\title{Braneworld Instantons}
\renewcommand{\thefootnote}{\fnsymbol{footnote}}
\author{ Ruth Gregory\footnote{R.A.W.Gregory@durham.ac.uk} and 
Antonio Padilla\footnote{Antonio.Padilla@durham.ac.uk}}
\address{Centre for Particle Theory, 
         Durham University, South Road, Durham, DH1 3LE, U.K.}
\date{\today}
\setlength{\footnotesep}{0.5\footnotesep}

\maketitle
\begin{abstract}

We derive the general equations of motion for a braneworld containing a
(localized) domain wall. Euclideanization gives the geometry for braneworld 
false vacuum decay.  We explicitly derive these instantons and compute 
the amplitude for false vacuum decay. We also show how to construct a toy 
ekpyrotic instanton. Finally, we comment on braneworlds with a compact spatial
direction and the adS soliton.

\end{abstract}
\pacs{PACS numbers: 04.50.+h, 11.25.Mj \hfill hep-th/0107108}

\renewcommand{\thefootnote}{\arabic{footnote}}

\section{Introduction}

The braneworld scenario \cite{RSh,KA}, \ie\ the idea that our observable
four-dimensional universe is somehow a localized `defect' or worldbrane in a
higher dimensional spacetime, has seen an almost inflationary-like expansion
of development and understanding recently, 
motivated in part by interest in unusual geometric
resolutions of the hierarchy problem~\cite{ADD,RS1,RS2}, but also 
by the ubiquity of branes in string and M-theory~\cite{LOSW}. The main feature 
distinguishing the braneworld from a standard Kaluza-Klein (KK) dimensional 
reduction is that physics is not averaged over these
extra dimensions, but strongly localized on this braneworld.
As such, it provides a compelling alternative to standard compactification
for theories in which extra dimensions are mandatory.

At the risk of oversimplification, the classes of models considered 
in nonstandard compactification can be split into roughly two categories:
those with `flat' extra dimensions (or nongravitating branes) such as the
original models of Arkani-Hamed et.\ al.\ (ADD)\cite{ADD}, and those with
warped extra dimensions (or self-gravitating branes) such as the
Randall-Sundrum (RS) models  \cite{RS1,RS2}. Generally, the former ADD
models have the braneworld sitting in a compact space 
of dimension two or higher with a standard gravitational KK reduction, 
whereas the RS models have only a single extra dimension -- a domain 
wall universe(s) living at the edge of anti-de Sitter (adS) spacetime,
with an exotic KK gravitational reduction \cite{EJS}. However, there
are variations on these themes which include features of both, such
as descriptions which consistently include gravity in the vacuum
ADD models \cite{CEG}, and higher co-dimension RS-style universes with 
both idealized \cite{GhSh}, as well as `thick brane' sources \cite{CK,NGE2}.

The key feature which makes the RS scenario so compelling and viable a
brane-{\it universe} model, is that is that gravity on
the domain wall is precisely Einstein gravity at low energies, \ie
\be \label{4deeq}
R_{\mu\nu} - {1\over2} R g_{\mu\nu} = 8\pi GT_{\mu\nu} 
\ee 
This result is of course perturbative \cite{RS2,GT}, and does not include
the effect of the short-range KK corrections. Also, strictly, it is only 
valid for a single brane universe -- the presence of a second wall introduces 
a radion, representing the distance between the branes, which modifies the
Einstein gravity to Brans-Dicke gravity \cite{GT,CGR,Ch}.
Non-perturbative results however, particularly understanding the effect
of the KK modes, are somewhat sparse. The simplest exact solutions 
(and those first explored) are the `zero-mode' solutions \cite{CG,CHR},
which take a four-dimensional Einstein solution on the braneworld
and keep it translationally invariant in the extra dimension,
which contains all the nontrivial adS behaviour. Such solutions
obviously obey the 4D Einstein equations (\ref{4deeq}), however,
they do not represent localized objects in a five dimensional sense,
and they tend to suffer from singularities at the adS horizon,
as well as instabilities, \cite{BSINS},  which indicate a fairly crucial
intervention of the KK modes in the gravitational field on the brane.
Perhaps the most interesting non-perturbative solution --
a black hole bound to the brane -- requires a five-dimensional
C-metric, which is so far unforthcoming. (See~\cite{EHM} for lower, 3+1,
dimensional analogues.)

Perhaps the most important non-perturbative solutions are the
cosmological solutions~\cite{BL,GCOS,COS,BCG,ACOS}, which
represent idealized brane FRW cosmologies. These also suffer from
singularities, although these are more of a cosmological variety; more
importantly however, they display departures from the 4D Einstein behaviour. 
The effect of the KK modes in this case is to introduce non-standard 
higher order terms in the Friedmann equation, first noted by Binetruy 
et.\ al.\ \cite{BL}, and later computed for completely arbitrary braneworlds
matching different adS spacetimes \cite{ACOS,BCG}.
Cosmological solutions also have an interesting link to string theory
via the adS/CFT correspondence (see e.g.\ \cite{GVS,CFT} for discussions).

Our interest is in non-perturbative solutions of a slightly different nature.
Recall that the key original observation of Rubakov and Shaposhnikov in
their domain wall universe set-up was in the interpretation
of zero-modes on the domain wall as confined particles living on the brane. 
For example, suppose we have a $\lambda \phi^4$ kink interacting
with an additional scalar, $\sigma$, via a potential of the form
\be\label{genpotl}
V(\phi,\sigma) = {\lambda\over4} (\phi^2-\eta^2)^2 + {{\tilde \lambda}\over4}
\sigma^4 + (\phi^2 - m^2) \sigma^2.
\ee
Then, the state $\langle \sigma \rangle=0$, while stable in the true vacuum
$\langle \phi\rangle=\pm\eta$, is tachyonic at the core of the wall; 
the scalar field $\sigma$ then develops a condensate on the core
of the wall $\sigma_0(z)$ (writing $z$ for the perpendicular coordinate).
One can therefore envisage this scalar as an inflaton --
perhaps stuck for a while in a false vacuum, and having a choice of
true vacua (the condensates) in which to roll. The above potential would
lead to a time-dependent slow roll problem, however, it also
raises the (asymptotically static) possibility of rolling to alternate
vacua in different asymptotic regions of the braneworld, reminiscent of
the tachyonic kink of non-BPS D-branes\cite{Sen}. Such a configuration --
a kink in the $\sigma$-field lying entirely within the core of the wall --
is quite well studied, at least in the absence of gravity, in the context
of nested topological defects in field theory \cite{Morris,Baz}, and this
particular configuration is known as a {\it domain ribbon} \cite{Morris}.

In a previous paper \cite{GP}, we explored the self-gravity of an idealized
domain ribbon - a `vortex', or codimension 2 delta-function source
nested inside the brane universe. These idealized sources could be included 
in the Israel junction conditions \cite{Israel} yielding a well-defined 
spacetime geometry. The spacetime turns out to be a general adS bulk
with a kinked boundary -- the kink being the location of the ribbon. This
set-up then describes nested domain walls in particular, a ``nested
RS scenario'' was introduced in \cite{GP}\ consisting of a Minkowski 
ribbon on an adS braneworld. Clearly however, by analytic continuation,
the domain ribbon solution and a braneworld instanton are closely related,
just as the domain wall and false vacuum decay instanton are 
closely related via the work of Coleman and de Luccia \cite{CDL}. 

In this paper, we give a detailed derivation of the domain ribbon spacetime,
including the possibility of non $Z_2$-symmetric braneworlds (section
\ref{sect:dr}). In section \ref{sect:solns} we derive some explicit solutions,
both to gain intuition for the domain ribbon spacetime as well as deriving
some useful canonical geometries for building instantons, which is the
topic of section \ref{sect:inst}. In particular, in section \ref{sect:inst},
as well as calculating the tunneling amplitude for false vacuum decay, 
we also show how to construct a toy `ekpyrotic instanton' a simplified form
of the brane peeling used in the ekpyrotic scenario \cite{EKP}.
Finally, in section \ref{sect:disc} we conclude and discuss braneworlds
with a compact spatial direction.

\section{The domain ribbon}\label{sect:dr}

Consider the gravitational field generated by a domain ribbon source.
In general, it will depend on only two
spacetime coordinates, $r$ and $z$ say, with $z$ roughly
representing the direction orthogonal to the domain wall 
(our brane universe), and $r$ the direction orthogonal to the domain
ribbon (or vortex) within our brane universe.
Schematically, the energy-momentum 
tensor of this source will have the following form:
\be
T_{ab} = \sigma h_{ab} {\delta(z) \over \sqrt{g_{zz}}}
+ \mu \gamma_{ab} {\delta(z) \delta(r) \over \sqrt{g_{zz}g_{rr}}}
\label{emtens}
\ee
where $h_{ab}$ is the induced metric on the brane universe, and $\gamma_{ab}$
the induced metric on the vortex. The symmetries of this energy-momentum
tensor mean that we can treat the vortex as a constant curvature spacetime. 
The most general metric consistent with
these symmetries can (generalizing \cite{BCG}) in $n$-dimensions
be reduced to the form
\be
ds^2 = A^{2\over(n-2)} d{\bf x}_\kappa^2 - e^{2\nu} A^{-{(n-3)\over(n-2)}} 
(dr^2 + dz^2)
\ee
where $d{\bf x}_\kappa^2$ represents the `unit' metric on a constant curvature
spacetime ($\kappa=0$ corresponds to an $(n-2)$-dimensional
Minkowski spacetime, $\kappa = \pm 1$ to $(n-2)$-dimensional
de-Sitter and anti-de Sitter spacetimes). $A$ and $\nu$ are functions
of $r$ and $z$ to be determined by the equations of motion. Here, 
the brane universe sits at $z = 0$, the vortex at $r = z = 0$. 
This is basically an analytic continuation of the cosmological metric
in~\cite{BCG}, where it is the time translation symmetry $\partial_t$ which is
broken, rather than $\partial_r$. The key result of that paper of relevance
here was to show that the conformal symmetry of the $t,z$ plane meant that
the gravity equations were completely integrable in the bulk, and the 
brane universe was simply a boundary ($T(\tau), Z(\tau)$) of that
bulk (identified with another boundary of another general bulk). The 
dynamical equations of the embedding of the boundary reduced to
pseudo-cosmological equations for $Z(\tau)$. We now briefly review this
argument in the context of the current problem, noting that it is
essentially analogous to a generalized Birkhoff theorem \cite{Birk}.

First of all, transform the $(r,z)$ coordinates to complex coordinates 
$(\omega, \bar{\omega})$ where $\omega=z+ir$, $\bar{\omega}=z-ir$, in which
the bulk equations of motion reduce to:
\bml \label{complextraj}
\bea 
\partial \bar{\partial} A &=& {1 \over 2}|\Lambda|A^{1 \over n-2}e^{2\nu} 
+ {(n- 2)(n-3)\over 4}\kappa A^{1 \over 2-n}e^{2\nu} \label{complextraja} \\
\partial \bar{\partial } \nu &=& {1 \over 4(n-2)}|\Lambda|A^{3-n \over n-2}
e^{2\nu} + {3-n\over 8}\kappa  A^{1-n\over n-2}e^{2\nu} \label{complextrajb} \\
\partial A \partial [\ln\partial A] &=& 2\partial \nu \partial A  
\label{complextrajc} \\
\bar{\partial} A \bar{\partial} [\ln\bar{\partial}A] &=& 
2\bar{\partial} \nu\bar{ \partial} A \label{complextrajd}
\eea
\eml 
where $\partial$ and $\bar{\partial}$ denote partial differentiation with
respect to $\omega$ and $\bar{\omega}$ respectively. For nonzero $\Lambda$ or
$\kappa$, equations (\ref{complextrajc}) and (\ref{complextrajd}) can be
integrated to give $e^{2\nu}=A^{\prime}f^{\prime}g^{\prime}$, where
$A = A \big( f(\omega)+g(\bar{\omega}) \big)$ with $f$ and $g$ being
arbitrary functions of the complex variables. The remaining equation
(\ref{complextraja}) for $A$ becomes an ODE.

Were the brane not present, we could use the fact that the metric depends only
on the combination $f+g$ to make a coordinate transformation in the bulk which
would give the metric in the familiar simple canonical form
\be\label{canbulk}
ds^2 = Z^2 d{\bf x}_\kappa^2 - h(Z) dR^2 - {dZ^2 \over h(Z)} 
\ee
where $d{\bf x}_\kappa^2$ is now a constant curvature Lorentzian spacetime,
and in general the function $h$ is
\be
h(Z) = k_n^2 Z^2 + \kappa - {c\over Z^{(n-3)}}
\ee
where we have encoded the cosmological constant in the form
\be
k_n^2 = -2{\Lambda\over(n-1)(n-2)}\label{lamkn}
\ee
If $c<0$, the metric becomes singular at
the adS horizon, $Z=0$. However, if $c>0$, the metric has the form of
a euclidean Schwarzschild-adS black hole times a constant curvature 
Minkowski geometry. The $\{R,Z\}$ plane closes off at the `euclidean
horizon', defined by $h(Z_+)=0$, where in order to avoid a conical 
singularity we must, in the usual fashion, identify $R$ as an 
angular coordinate with periodicity
$\Delta R = 4\pi/h'(Z_+)$. The geometry of the $\{R,Z\}$ sections is
therefore (up to an asymptotic adS warp factor) the familiar ``cigar'' 
with a smooth tip at $Z_+$.  This is the adS soliton considered
by Horowitz and Myers \cite{HM}.

The addition of the brane however requires that the Israel conditions be 
satisfied at $z=0$ in the original coordinates. These 
turn out to have a scaling symmetry $\omega \to W(\omega)$, 
${\bar\omega}\to W({\bar\omega})$, hence we have a residual gauge 
freedom which allows us to specify one of $f$ or $g$ as we wish. 
The net result is that our brane becomes some boundary of the bulk 
metric (\ref{canbulk}) identified with the boundary of some other 
general bulk metric. The vortex (or ribbon), in these coordinates, 
becomes a kink 
on this boundary as we shall see.  Introducing the affine parameter $\zeta$ 
which parametrizes geodesics on the brane normal to the vortex, 
the brane is now  given by the section $({\bf x}^\mu , R(\zeta), Z(\zeta))$ 
of the general bulk metric. 

In an exactly analogous procedure to \cite{BCG}, we consider the Israel 
equations  for the jump in extrinsic curvature across the brane, as well 
as the normal component of the Einstein equations, and thus obtain the
equations of motion for the source:
\bml \label{srctraject}
\bea
Z^{\prime 2} &=&\bar{h}-\sigma_n^2Z^2-\left({ \Delta h  \over4\sigma_nZ}  
\right)^2 \label{srctrajecta} \\
Z^{\prime \prime} &=& {\bar{h^{\prime}} \over 2}-\sigma_n^2Z 
-{\mu_n \over2}\sigma_nZ\delta(\zeta) 
+ {1 \over Z}\left({ \Delta h \over4\sigma_nZ}  \right)^2 \nonumber \\  
&-&{\Delta h^{\prime}\Delta h\over(4 \sigma_n Z)^2}+{\mu_n\delta(\zeta)\over2
\sigma_nZ }\left({ \Delta h\over4\sigma_nZ}\right)^2 \label{srctrajectb} \\
\overline{\epsilon h R^{\prime}} &=& \sigma_nZ \label{srctrajectc}
\eea
\eml
These equations are extremely general and therefore have some of their content
disguised by notation that should be defined. The brane and vortex tensions
have for convenience been normalized to
\be\label{rescalesigmu}
\sigma_n = {8\pi G_n \sigma\over 2(n-2)} \qquad ; \qquad\mu_n = 8\pi G_n\mu
\ee
respectively. Recall that we consider the brane to be the boundary 
$({\bf x} ^\mu, R(\zeta), Z(\zeta))$ of two distinct spacetimes: 
the $(+)$ spacetime in $z>0$ and the $(-)$ spacetime in $z<0$. 
Quantities that are intrinsic to a given bulk spacetime acquire a suffix
associated with that spacetime. For example $h_+$ corresponds to 
$h$ in the $(+)$ spacetime. Furthermore, for any such quantity $Q$, we define
the average and difference across the brane as follows:
\bml
\bea
\bar{Q} &=& {1\over2}(Q_++Q_-) \\
\Delta{Q} &=& Q_+-Q_-
\eea
\eml
Finally, the quantity $\epsilon$ in (\ref{srctrajectc}) is related to the sign
of the the normal to the boundary of the bulk spacetime, the boundary of 
course being the brane. This normal is an `inward pointing' normal, that is, 
its (contravariant) components point \emph{into} the bulk which is being
retained on the $(+)$ side of the boundary, and \emph{away} from the retained
bulk on the $(-)$ side of the boundary.
For example,  the boundary to the $(+)$ spacetime has a normal pointing 
inwards given by 
\be
n_a=\epsilon_+({\bf 0}, -Z^{\prime}, R^{\prime})
\ee
where $\epsilon_+$ can take values of $\pm 1$.

For simplicity, we will now assume our
brane universe is $Z_2$-symmetric (\ie\ spacetime is reflection symmetric
around the wall). This has the effect that any intrinsic bulk quantity $Q$ 
loses its suffix and so $\bar{Q} \to Q$ and $\Delta Q \to 0$. 
We also assume for now that the integration constant, $c$, vanishes.
Rewriting equation (\ref{srctraject}) for the trajectory of the source 
we obtain the $Z_2$-symmetric equations of motion: 
\bml\label{traject}\bea
Z^{\prime2}(\zeta) &=& \left ( k_n^2 - \sigma^2_n \right )
Z^2 + \kappa \label{trajecta} \\
Z''(\zeta) &=& \left ( k_n^2 - \sigma^2_n \right )
Z - {\mu_n\over 2} \sigma_n Z \delta(\zeta) \label{trajectb} \\
R'(\zeta) &=& {\sigma_n Z\over (k_n^2 Z^2 + \kappa)} \label{trajectc} 
\eea\eml

For example, the Randall-Sundrum domain wall (in $n$-dimensions) is given
by setting $\kappa = \mu = 0$ (flat, no vortex) and $\sigma_n
= k_n$.  The bulk metric is then
\be
ds^2 = Z^2 (dt^2 - dx^2_i -k^2_n dR^2) - {dZ^2\over k_n^2Z^2} 
\ee
and we have the solution $Z=Z_0$ a constant, and $kR = \zeta/Z_0$.
Letting $Z_0=1$, and $Z = e^{-k_nz}$ gives the usual RS coordinates.
Replacing the Minkowski metric (in brackets) by an arbitrary 
$(n-1)$-dimensional metric gives the usual relation between 
Newton's constant in $n$ and $n-1$ dimensions for the RS universe:
\be\label{newton}
G_{n-1} = {(n-3)\over2} k_nG_n = {(n-3)\over2} \sigma_nG_n
\ee
a relationship confirmed by the perturbative analysis
of~\cite{RS2,GT} for the critical brane, and \cite{GS} for the 
de Sitter brane.

Before turning to the instanton solutions, we will remark on a few 
straightforward domain ribbon solutions in order to gain an understanding 
of the geometrical effect of the ribbon.

\section{Specific solutions}\label{sect:solns}

In this section we examine the solutions to (\ref{traject}), exploring their 
qualitative features as well as some useful illustrative special cases. To 
start, note that the $Z$-equation (\ref{trajecta}) can be integrated (away 
from the vortex) to give
\be\label{genzsolns}
Z = \cases{ {1\over 2\sqrt{a}} \left [
e^{\pm\sqrt{a}(\zeta-\zeta_0)} - \kappa e^{\mp \sqrt{a}(\zeta-\zeta_0)} 
\right] & $a>0$\cr
Z_0 \pm \kappa \zeta & $a=0$, $\kappa = 0,1$ \cr
{1\over \sqrt{|a|}} \cos\pm\sqrt{|a|}(\zeta - \zeta_0) & $a<0$, 
$\kappa = 1$ only \cr}
\ee 
where $a = k_n^2 - \sigma^2_n$, which is zero for a critical wall, and is
positive (negative) for a sub (super) critical wall respectively.
In the absence of the vortex, a critical wall is one with a
Minkowski induced metric, and is the original RS scenario \cite{RS1,RS2}, a
supercritical wall is one which has a de-Sitter induced metric, and
can be regarded as an inflating cosmology \cite{COS}, whereas
the subcritical wall has an adS induced metric, and has only recently
been considered from the phenomenological point of view \cite{KR}.
In addition, also away from the vortex, we can use (\ref{trajectc}) and
the root of (\ref{trajecta}) to obtain an equation for $R(Z)$ which is
easily integrated to yield
\be\label{Rtraject}
2k_n (R - R_0) =  \pm \cases{\ln\left (1 + k_n^2 Z^2 \right) &
$\kappa = 1$, $a=0$ \cr
\ln\left | {k_n\sqrt{1+aZ^2} - \sigma_n\over k_n\sqrt{1+aZ^2}+\sigma_n}
\right |  & $\kappa = 1$, $a\neq0$\cr
-{2\sigma_n\over k_n \sqrt{a} Z} & $\kappa = 0$, $a>0$ \cr
2 \tan^{-1}\left ( {k_n\sqrt{1 + aZ^2} \over\sigma_n} \right ) &
$\kappa = -1$, $a>0$.\cr}
\ee
where the choice of signs refers to the sign of $Z'(\zeta)$.
Note that these trajectories are invariant under euclideanization of
the metric, therefore instanton trajectories will also have this form.

In order to see how these trajectories embed into the bulk of adS
spacetime, it is useful to transform into conformal coordinates,
$\{{\tilde t}, {\tilde{\bf x}}, u\}$ in which the metric is conformally
flat:
\be\label{confmet}
ds^2 = {1\over k_n^2u^2} \left [ d{\tilde t}^2 - d{\tilde {\bf x}}^2 - du^2
\right ]
\ee
For the $\kappa=1$ spacetimes which we will need to construct the braneworld
instantons, this requires the bulk coordinate transformation
\bml\label{bulkct}\bea
k_n u &=& e^{k_nR} / \sqrt{1+k_n^2 Z^2} \\
\left ( {\tilde t}, {\tilde {\bf x}}\right ) &=& k_n u Z 
\left ( \sinh t , \cosh t {\bf n}_{n-2} \right )
\eea\eml
(where ${\bf n}_{n-2}$ is the unit vector in $(n-2)$ dimensions). 
Under such a transformation the trajectories $R(Z)$ in (\ref{Rtraject})
become in general of the form
\be\label{hyper}
(u \mp u_0)^2 + {\tilde {\bf x}}^2 - t^2 = u_1^2
\ee
for nonzero $a$, where
\be
u_0 = {\sigma_n \over k_n} u_1 
= {\sigma_n \over k_n} {e^{k_nR_0}\over |a|^{1/2}}
\ee
\ie\ the braneworlds (\ref{hyper}) have the form of hyperboloids (or
spheres in the euclidean section) in the conformal metric (\ref{confmet}).
\begin{figure}
\begin{center}
\epsfig{file=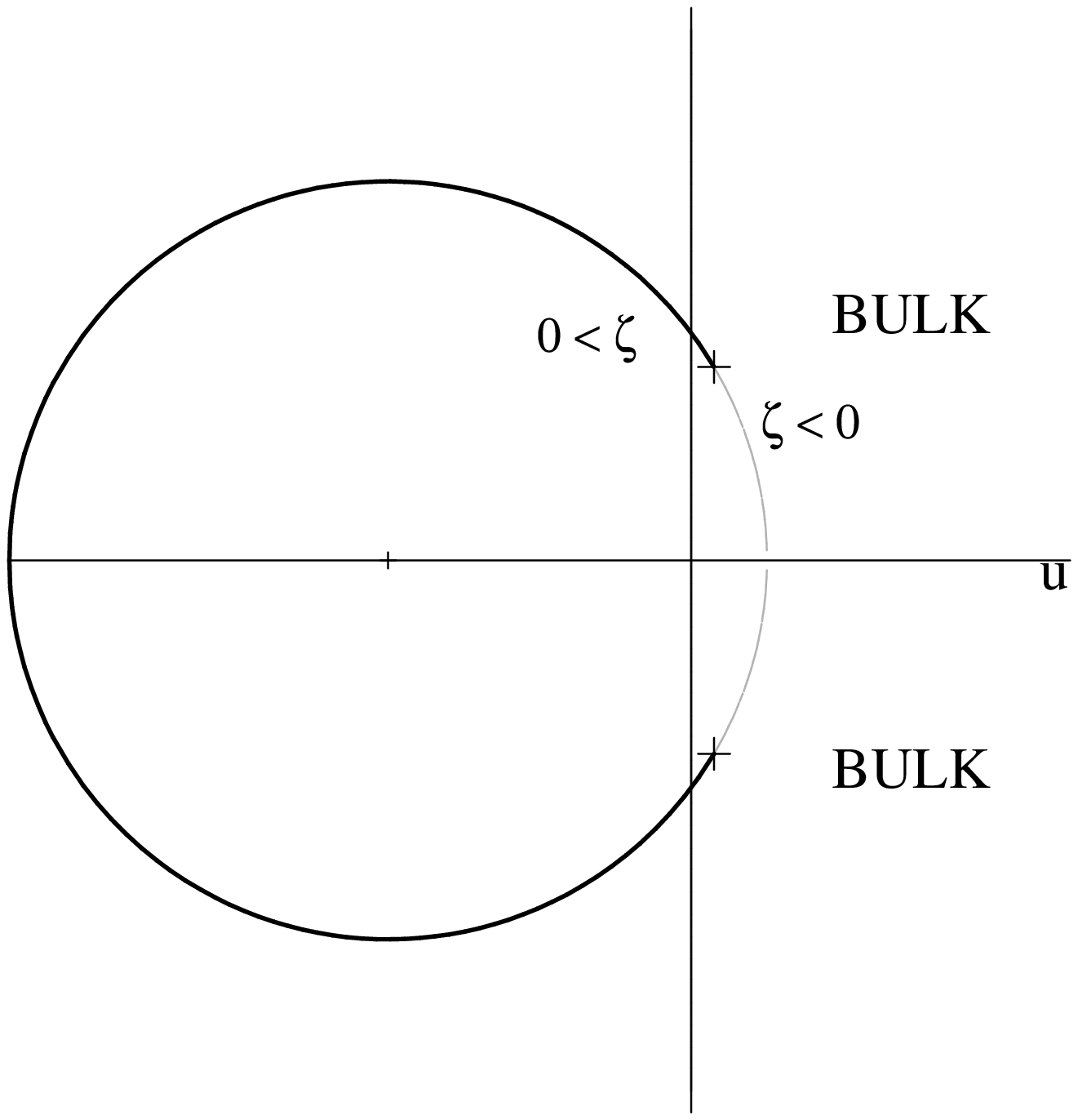,width=6.8cm} \qquad
\epsfig{file=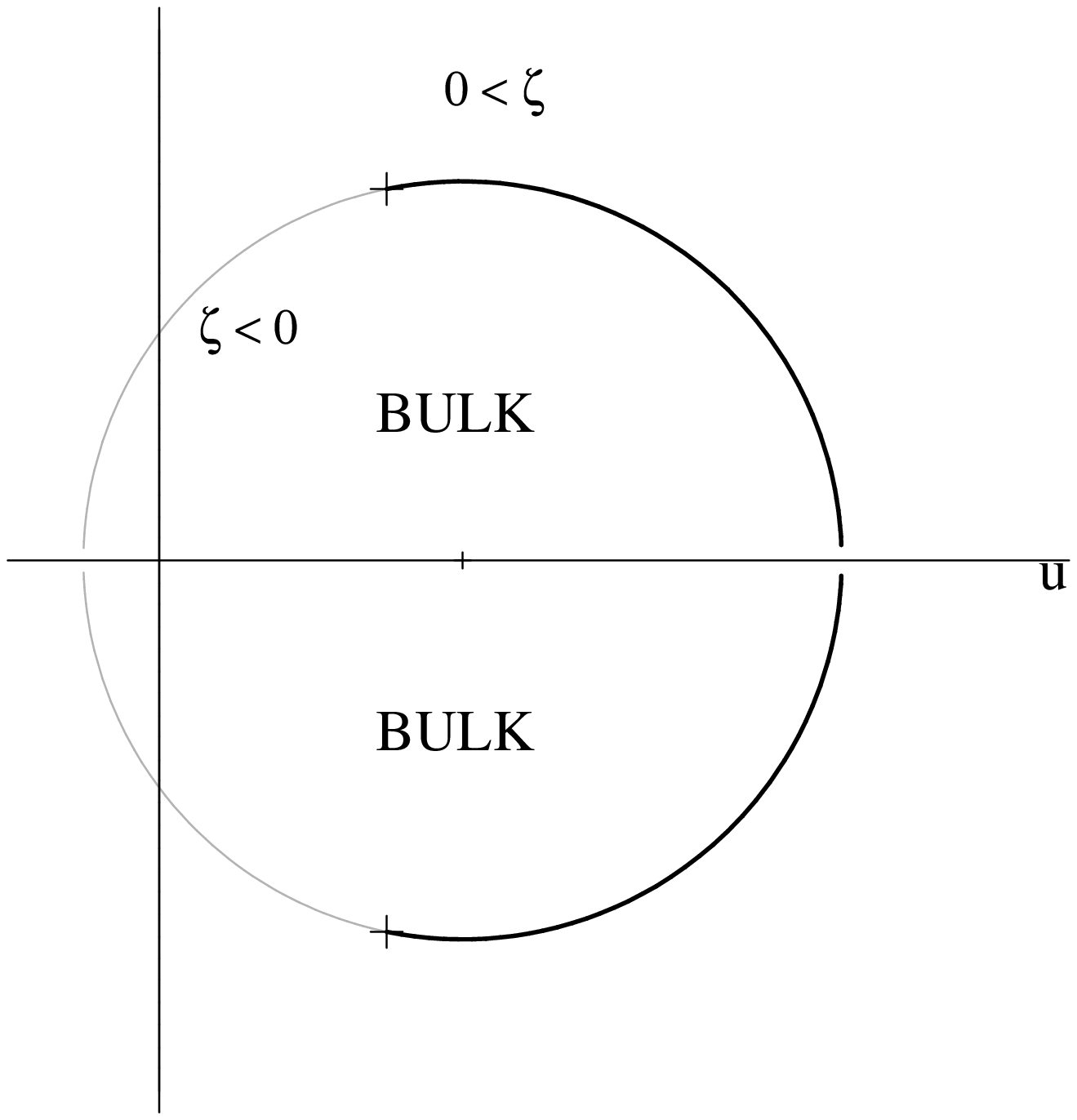,width=6.8cm} \\
(a)  Subcritical wall centered on $-u_0$. \hskip 2cm 
(b) Subcritical wall centered on $+u_0$. \\ ~~\\
\epsfig{file=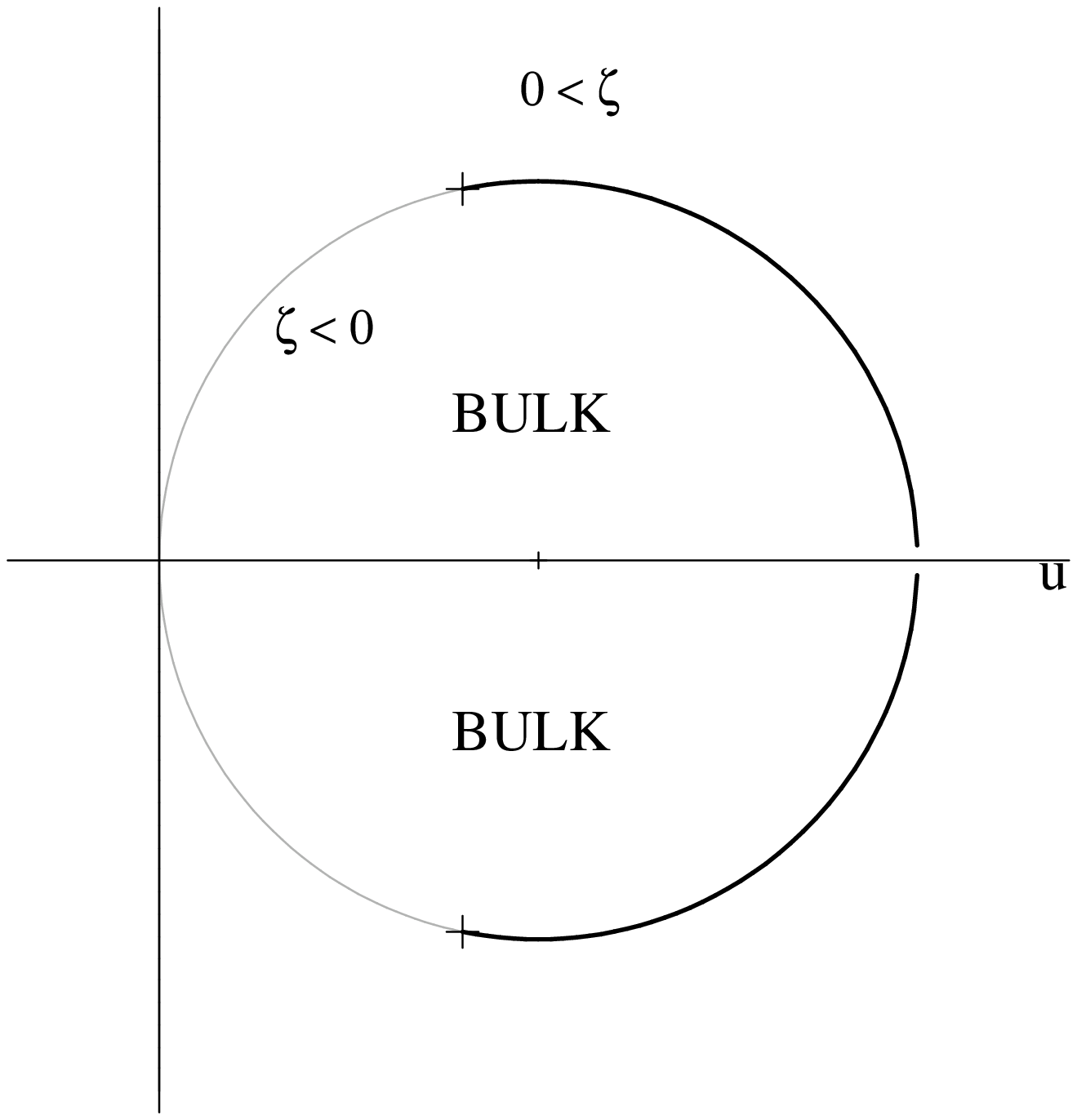,width=6.8cm}   \qquad
\epsfig{file=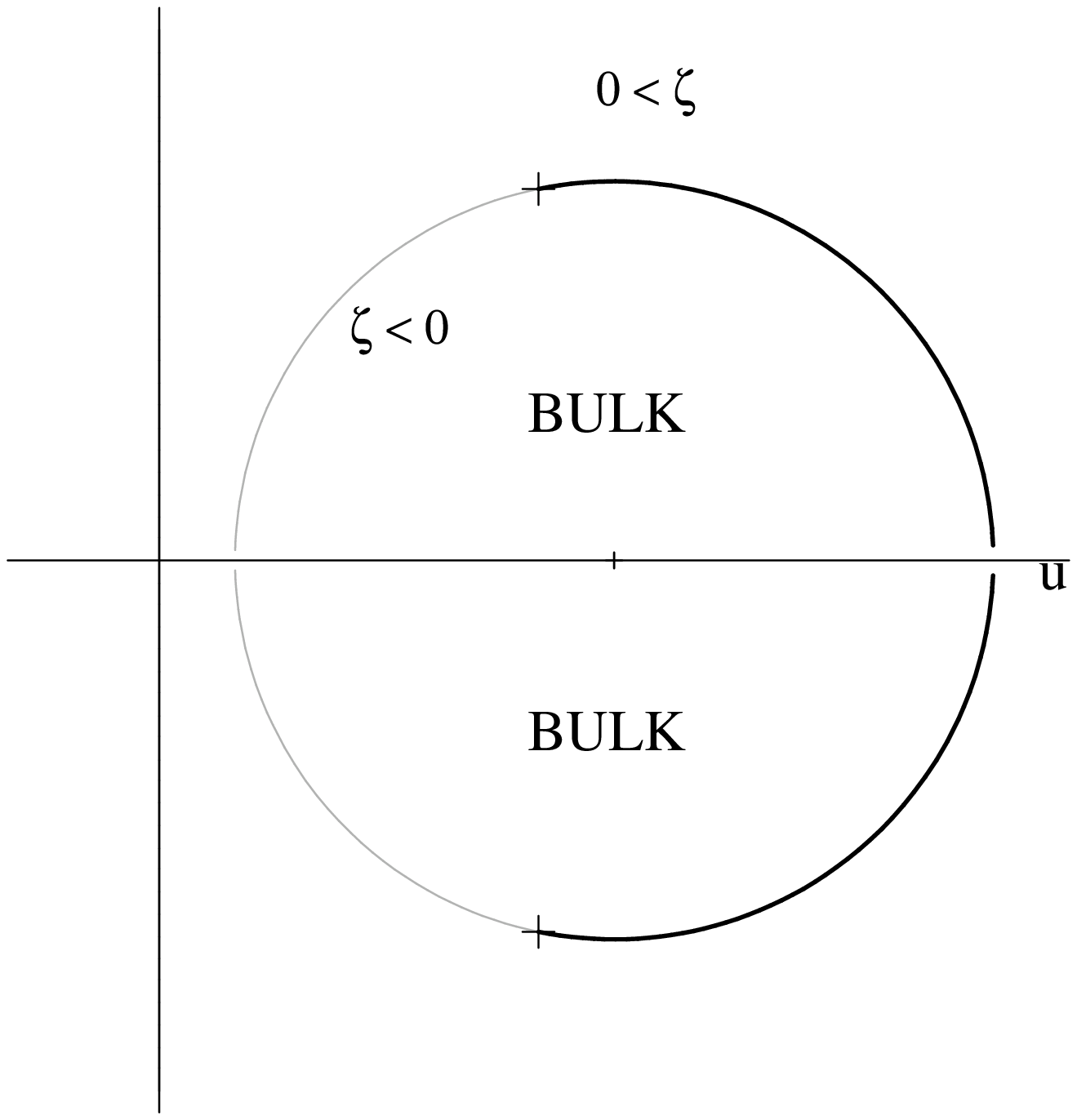,width=6.8cm} \\
(c) Critical wall. \hskip 5cm (d) Supercritical wall. \hskip 2cm\\
\end{center}
\caption{Wall trajectories given by equation \ref{hyper}, in 
euclidean signature. In each case the 
location of the bulk spacetime is indicated. In
addition we have the simple critical wall trajectory given by
$u=\text{const}$. Here the bulk lies to the right of the wall.}
\label{fig:walltraj}
\end{figure}

For subcritical walls, \ie\ $\sigma_n < k_n$, $u_0<u_1$, and both branches of
the hyperboloid (\ref{hyper}) are allowed, which both intersect 
the adS boundary -- see figure \ref{fig:walltraj}(a,b). An analysis of the
normals to the braneworld shows that for a positive tension $Z_2$-symmetric 
braneworld, the upper root $Z'>0$ corresponds to keeping the interior of
the hyperboloid, whereas for $Z'<0$ the exterior is kept.  Supercritical
walls on the other hand have only the upper root for $u_0$, and as $u_0>u_1$
in a euclidean signature they represent spheres which are entirely 
contained within the adS spacetime, the interior of the hyperboloid (or
sphere) being kept for a positive tension braneworld (see figure 
\ref{fig:walltraj}(d)). For a critical wall, $\sigma_n=k_n$ there are once
again two possible trajectories, one having the form of (\ref{hyper})
but with $u_0=u_1 = e^{k_nR_0}/2k_n$ (figure \ref{fig:walltraj}(c)), and 
the other having $u=$ const.\ -- the RS braneworld.

To put a vortex on the braneworld, 
we require solutions with nonzero $\mu_n$, and hence a discontinuity in $Z'$.
To achieve this, we simply patch together different branches of the 
solutions (\ref{genzsolns}) for $\zeta>0$ and $\zeta<0$. 
We immediately see that critical and supercritical walls can
only support a vortex if $\kappa=1$, \ie\ if the induced metric on the vortex
itself is a de-Sitter universe. A subcritical wall on the other hand can
support all induced geometries on the vortex. 
Defining $k_{n-1}^2 = |a|$, these trajectories are
\be\label{zsolns}
Z = \cases{ {1\over 2k_{n-1}} \left [
e^{\alpha-k_{n-1}|\zeta|} - \kappa e^{k_{n-1}|\zeta|-\alpha} 
\right] & subcritical wall \cr
Z_0 - |\zeta|& critical wall ($\kappa=1$) \cr
{1\over k_{n-1}} \cos \left (k_{n-1}|\zeta| + \beta\right) & supercritical wall 
($\kappa = 1$) \cr}
\ee 
where
\be\label{muvalues}
\mu_n = \cases{ {4k_{n-1}\over\sigma_n} \left [
{e^{2\alpha} + \kappa \over
e^{2\alpha} -\kappa } \right ] & $\kappa=\pm1,0$
subcritical wall \cr
{4\over k_n Z_0} & critical wall \cr
{4k_{n-1}\over\sigma_n} \tan \beta & supercritical wall \cr}
\ee
respectively. 

A particularly useful example, for the purposes of illustrating
the geometry of the ribbon, is to consider a vacuum bulk spacetime, 
which will obviously represent a supercritical braneworld.
This removes any question of the warping of the bulk due to the
cosmological constant, and shows particularly clearly the similarities
and differences between the ribbon spacetime, an isolated vortex, and
a domain wall in $(n-1)$-dimensional
de-Sitter spacetime (the supercritical braneworld). 

The case where there is no bulk cosmological constant has $k_n=0$, hence
$k_{n-1} = \sigma_n$, and a pure domain wall universe would be a 
hyperboloid \cite{CGG} -- an accelerating bubble of proper 
radius $\sigma_n^{-1}$ in Minkowski spacetime, for which $\kappa=1$.  
For future comparison, we also note that a pure $\delta$-function
vortex solution in a vacuum spacetime has a conical deficit metric
\be\label{conic}
ds^2 = dt^2 - d{\bf x}^2 - d\rho^2 - 
\left (1-{\Delta\theta\over2\pi}\right)^2 \rho^2d\theta^2
\ee
where $\Delta\theta \simeq \mu_n$ for small $\mu_n$\cite{VIL}.

Reading off the domain ribbon trajectory from (\ref{zsolns}) gives
in $(R,Z)$ space
\bml\label{fwtraj}\bea
Z &=& {1\over \sigma_n \sqrt{16+\mu_n^2}} \left [ 4 \cos(\sigma_n\zeta) 
- \mu_n \sin(\sigma_n|\zeta|)\right ] \\
R &=& {1\over \sigma_n \sqrt{16+\mu_n^2}} \left [ 4 \sin(\sigma_n\zeta)\pm
\mu_n [ \cos(\sigma_n\zeta) -1] \right]
\eea\eml
preserving the region $Z<Z(\zeta)$ of the bulk:
\be
ds^2 = Z^2 \left[ dt^2 - \cosh^2\! t\ d\Omega_{n-3}^2 \right]
- dZ^2 - dR^2
\ee
This is of course simply a coordinate transformation of 
Minkowski spacetime, with the appropriate limit of (\ref{bulkct}) being
$({\tilde t}, {\tilde{\bf x}}) = 
(Z\sinh t, Z{\bf n}\cosh t)$.
Transforming into Minkowski coordinates therefore, we find that
the vacuum braneworld domain ribbon is given by two copies of the interior of 
the sliced hyperboloid
\be
{\tilde{\bf x}}^2 - {\tilde t}^2 + \left (|R|+{\mu_n\over\sigma_n 
\sqrt{16+\mu_n^2}} \right )^2 = {1\over\sigma_n^2}
\ee

\begin{figure}
\centerline{\epsfig{file=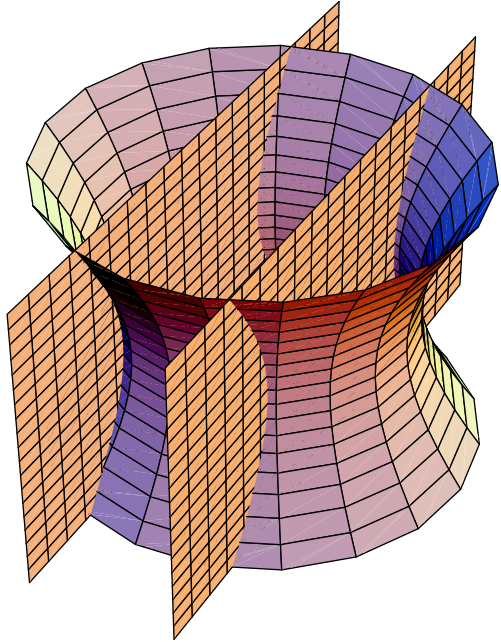,width=8cm}}
\vskip 5mm
\caption{Constructing a domain ribbon on a vacuum domain wall. The hyperboloid 
interior has a slice of thickness $2\mu_n/\sigma_n \sqrt{16+\mu_n^2}$ 
removed from it, and is re-identified. The full spacetime consists of
a second copy identified across the hyperboloid.}
\label{fig:vacdw1}
\end{figure}

\begin{figure}
\centerline{\epsfig{file=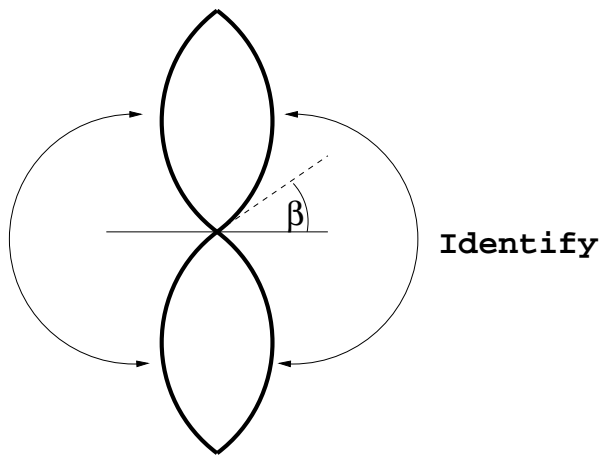,width=8cm}}
\vskip 5mm
\caption{Taking a constant time slice through the vacuum domain wall
plus vortex spacetime shows how the deficit angle is built up.}
\label{fig:vacdw2}
\end{figure}

If $\mu_n=0$ this is clearly the standard domain wall hyperboloid, however,
for $\mu_n>0$ this now represents a hyperboloid which has had a slice
of width $2\mu_n/\sigma_n \sqrt{16+\mu_n^2}$ removed from it
(see figure \ref{fig:vacdw1}). 
This corresponds rather well with the intuitive notion
that walls are obtained by slicing and gluing spacetimes, and looking
at a constant time slice (figure \ref{fig:vacdw2}) we also see 
how the domain ribbon 
looks like a vortex, with the identifications giving rise to a conical
deficit angle in terms of the overall $n$-dimensional spacetime. A
crucial difference here however appears to be that we can have an arbitrarily
large energy per unit length of this vortex, as we simply cut out more
and more of the hyperboloid. In terms of the conical deficit, we find
\be
\Delta \theta = 4 \tan^{-1} \mu_n/4
\ee
and therefore the deficit angle approaches $2\pi$ only as $\mu_n$ approaches
infinity! Contrast this with the spacetime of a pure vortex, (\ref{conic}), 
in which the
deficit angle approaches $2\pi$ as $\mu_n \simeq 1$ \cite{LG}, clearly the 
ribbon is not behaving as a vortex for large $\mu$. On the other hand, a 
domain wall has the effect of compactifying its spatial sections (the interior
of the hyperboloid) and the transverse dimension only shrinks to zero size as
the tension of the wall becomes infinite. Therefore in this sense, the ribbon
spacetime really does behave as a domain wall.

We can also write down the induced metric on the domain wall
brane-universe:
\be
ds_{n-1}^2 = {[4\cos\sigma_n\zeta - \mu_n \sin \sigma_n |\zeta|]^2\over
\sigma_n^2 (16+\mu_n^2)} \left [ dt^2 - \cosh^2t d\Omega^2_{n-3} \right]
-d\zeta^2
\ee
which is the metric of an $(n-2)$-dimensional domain wall in an
$(n-1)$-dimensional de-Sitter universe of tension
\be
G_{n-1} T = {(n-3) \sigma_nG_n\over 2} \mu
\ee
\ie\ with the identification (\ref{newton}) (using $\sigma_n$,
rather than $k_n$ which is of course zero in this case) $T\equiv \mu$.

\begin{figure}
\centerline{\epsfig{file=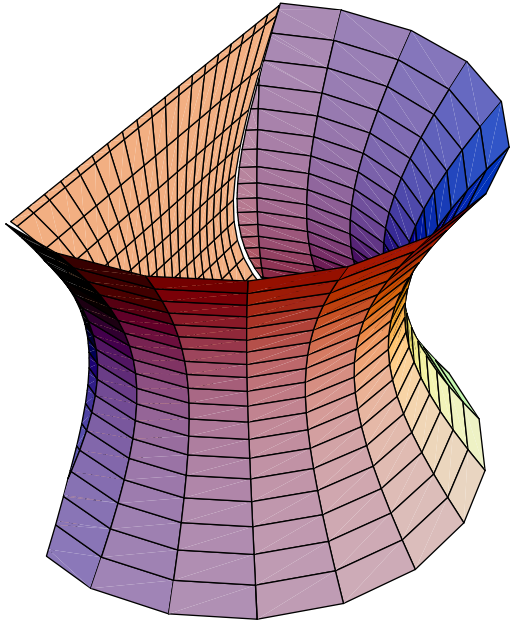,width=8cm}}
\vskip 5mm
\caption{A representation of the domain ribbon in a critical RS
universe. }
\label{fig:rsdw}
\end{figure}

Having constructed this symmetric vacuum domain ribbon spacetime,
we now see the general principle involved in having a domain ribbon.
Whereas the braneworld consists of two segments of adS (or vacuum/dS)
spacetime glued across a boundary, the domain ribbon consists of two
copies of an adS spacetime with a kinked boundary (which could itself
be viewed as two copies of an adS bulk glued together across a tensionless
boundary making a kink) identified together. Therefore a domain ribbon
on a critical wall in conformal coordinates is obviously the critical
hyperboloid sliced by a critical flat RS wall (see figure \ref{fig:rsdw}).
Once more, as $\mu$ increases, more and more of the hyperboloid
is removed, with the spacetime `disappearing' only as
$\mu\to\infty$.

We can also construct a range of nested Randall-Sundrum type configurations,
\ie\ flat ribbon geometries on adS braneworld trajectories within the
bulk adS spacetime.
From (\ref{Rtraject}) we see that a flat ribbon geometry
requires a subcritical wall, which in conformal coordinates has the form
\be
{\tilde x}^1 = \pm {\sigma_n\over k_{n-1}}(u-u_0)
\ee
For example, the original RSI domain ribbon would have the form
(taking the universal covering) of a sawtooth trajectory:
\be
R = 2j (u_1-u_0) {\sigma_n\over k_{n-1}}
\pm {\sigma_n\over k_{n-1}} (u-u_0) \;\;\; j\in Z
\ee
with alternate positive and negative tension ribbons. Similarly, one
can construct a nested GRS scenario \cite{GRS}\ :
\be
u = \cases{ u_0 - {k_{n-1} |R|\over\sigma_n} & $|R| 
< {\sigma_n\over k_{n-1}} (u_1 - u_0)$ \cr
u_1 & otherwise \cr}
\ee
although whether this displays a quasilocal graviton is not so easy to 
determine. 

In principle one can construct all sorts of nested multibrane scenarios,
however one suspects that the more baroque the model, the less stable
or useful it is likely to be.

\section{Instantons and tunneling on the brane}\label{sect:inst}

Traditionally, instantons correspond to classical euclidean solutions to the
equations of motion. In many cases, they represent a quantum tunneling from a
metastable false vacuum to a true vacuum. In ~\cite{CDL}, Coleman and
de Luccia discussed the effect of gravity on these decays. Such processes,
of course have direct relevance for cosmology, as they correspond to
a first order phase transition, and hence a
dramatic change in the structure of our universe.

In ~\cite{CDL}, the authors evaluated the probability of  
nucleation of a true vacuum
bubble in a false vacuum background. They focussed on two particular
configurations: a flat bubble spacetime in a de Sitter false vacuum;
and an adS bubble spacetime in a flat false vacuum. This was before
the idea of large extra dimensions was fashionable, so the analysis
was done in just the usual four dimensions.

We now have the tools to develop these ideas in a brane world set
up. To replicate the configurations of ~\cite{CDL}, we just have to
patch together our wall trajectories in the right way. Recall that
these trajectories are given by equation (\ref{hyper}), along with the
critical wall solution,  $u=\text{const}$. In euclidean signature, 
the former are shaped like spheres and were illustrated in figure 
\ref{fig:walltraj}. However, when patching these solutions together, 
we should be aware of a slight subtlety. In equation (\ref{trajectb}), 
the $\mu_n \sigma_n \delta(\zeta)$ term does not make sense if we have 
walls of different type either side of the vortex. Suppose we have a 
wall of tension $\sigma_n^+$ in $\zeta>0$, and $\sigma_n^-$ in $\zeta<0$, 
we must then modify equation (\ref{trajectb}) by replacing $\sigma_n$ 
with $\bar{\sigma}_n$, where 
\be
\bar{\sigma}_n={\sigma_n^+  + \sigma_n^- \over 2}
\ee
It is easy to see that this is the right thing to do. Regard the
vortex as a thin wall limit of some even energy
distribution. Mathematically, this corresponds to $\mu_n \delta(\zeta)$
being the limit of some even function $\mu_nf(\zeta)$. The weight of the
distribution is the same on either side of $\zeta=0$, so we pick up
the average of the wall tensions.

We are now in a position to reproduce the work of ~\cite{CDL} in our
higher dimensional environment. Let us consider first the decay of a
de Sitter false vacuum, and the nucleation of a flat bubble spacetime.

\subsection{Nucleation of a flat bubble spacetime in a de Sitter false
vacuum}

We now describe the brane world analogue of the nucleation of 
a flat bubble spacetime in a
de Sitter false vacuum. The de Sitter false vacuum is given
by a supercritical wall of tension $\sigma_n^{dS}>k_n$ 
with no vortex (see figure \ref{fig:walltraj}(d)). This
metastable state decays into a ``bounce'' configuration given by a
critical wall (tension $\sigma_n^{flat}=k_n$) 
patched on to a supercritical wall
(tension $\sigma_n^{dS}>k_n$). If we are to avoid generating an
unphysical negative tension vortex we must patch together trajectories
in the following way:
\be
Z=\cases{ {1 \over k_{n-1}^{dS}}\cos(k_{n-1}^{dS}\zeta-\zeta_0)
 &  $\zeta>0$ \cr
\zeta+{1 \over k_{n-1}^{dS}}\cos\zeta_0 &  $\zeta<0$  \cr
}
\ee
where $(k_{n-1}^{dS})^2=(\sigma_n^{dS})^2-k_n^2$. The vortex tension
$\mu$, 
is related to the constant $\zeta_0$ in the following way:
\be \label{matchingcond}
{\mu_n \bar{\sigma}_n \over 2  k_{n-1}^{dS}}=\sec\zeta_0-\tan\zeta_0
\ee

\begin{figure}
\begin{center}
\includegraphics[width=8cm]{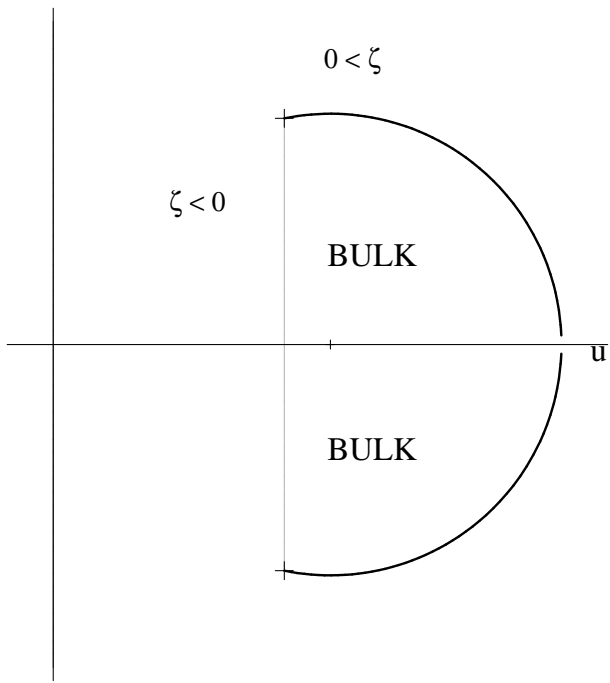}
\vskip 5mm
\caption{An example of a critical-supercritical wall ``bounce'' solution. 
This looks
like a flat bubble spacetime has nucleated on a de Sitter wall.} 
\label{fig:bounce1}
\end{center}
\end{figure}

It is useful to have a geometrical picture of this bounce solution. We
just patch together the $u=$ const.\ critical wall trajectory and the
supercritical wall trajectory given by figure \ref{fig:walltraj}(d) to get
figure \ref{fig:bounce1}. Note that we have two copies of the bulk
spacetime because we imposed $Z_2$-symmetry across the walls.

It is natural to calculate the probability, $\mathcal{P}$, that this 
flat bubble spacetime does indeed nucleate on the de Sitter wall. 
\be
{\mathcal{P}}  \propto e^{-B}
\ee
where $B$ is the difference between the euclidean actions of the bounce
solution and the false vacuum solution, that is:
\be
B=S_{bounce}-S_{false}
\ee
Given our geometrical picture it is straightforward to write down an
expression for the bounce action:
\begin{equation}
S_{bounce}=S_{bulk}+S_{flat}+S_{dS}+S_{vortex}
\end{equation}
where the contribution from the bulk, critical wall (flat),
supercritical wall (de Sitter), and vortex are as follows
\bml \label{bounceaction} \begin{eqnarray} 
S_{bulk} &=& -\frac{1}{16\pi G_{n}} \int_{bulk}
d^{n}x\sqrt{g}(R-2\Lambda) \label{bounceactiona} \\
S_{flat}&=& -\frac{1}{16\pi G_{n}}\int_{flat} d^{n-1}x
\sqrt{h}(-2\Delta K-4(n-2)\sigma_n^{flat})   \label{bounceactionb} \\
S_{dS} &=&  -\frac{1}{16\pi G_{n}}\int_{dS} d^{n-1}x
\sqrt{h}(-2\Delta K-4(n-2)\sigma_n^{dS})  \label{bounceactiond} \\
S_{vortex} &=& \mu \int_{vortex} d^{n-2}x \sqrt{\gamma}=
-\frac{1}{16\pi G_{n}}\int_{vortex} d^{n-2}x \sqrt{\gamma}(-2\mu_{n})  
\label{bounceactione}
\end{eqnarray}
\eml
Note that $\Delta K$, the jump in the trace of the extrinsic curvature 
across the wall, contains the Gibbons-Hawking boundary term \cite{GH} for 
each side of the wall, and  $h_{ab}$, $\gamma_{\mu\nu}$ are the induced 
metrics on the wall and vortex respectively. We should point out 
that due to the presence of the vortex, there is a delta function in 
the extrinsic curvature that exactly cancels off the contribution 
of $S_{vortex}$.
The expression for $S_{false}$ is similar except that there is no flat
wall or vortex contribution, and no delta function in the extrinsic 
curvature. After some calculation (see Appendix), we find that 
our probability term, $B$ is given by:
\be\label{probterm}
B={4k_n^{2-n}\Omega_{n-2} \over 16\pi G_n} \left( I_n-\left({1 \over
n-1} \right) \left({k_n\cos\zeta_0 \over
k_{n-1}^{dS}}\right)^{n-1} \right)
\ee
where $\Omega_{n-2}$ is the volume of an $(n-2)$ sphere and the integral
$I_n$ is given by:
\be \label{In}
I_n=\int_{u_0-u_1}^{u_c}  du \quad  \left ( u_0{[\rho(u)]^{n-3} 
\over u^{n-1}}-{[\rho(u)]^{n-1} \over u^n} \right )
\ee
where
\bml\label{dswallrelations}\bea
u_0 &=& {\sigma_n^{dS} \over (k_{n-1}^{dS})^2}(\sigma_n^{dS} +
k_n\sin\zeta_0)u_c \label{dswallrela} \\
u_1 &=& {k_n \over \sigma_n^{dS}}u_0  \label{dswallrelb}\\
\rho(u) &=& \sqrt{u_1^2-(u-u_0)^2} \label{dswallrelc} 
\eea\eml
Note that $u_c$ is an arbitrary constant so we are free to choose it as
we please (think of the flat wall as being at $u=u_c$). This integral is
non trivial and although we can in principle solve it for any integer
$n$ it would not be instructive to do so. Instead, we will restrict
our attention to the case where $n=5$. This means that our braneworld
is four dimensional, so comparisons with ~\cite{CDL} are more
natural. Given that $\Omega_3=2\pi^2$, we find that:
\be
B = {8\pi^2k_5^{-3} \over 16\pi G_5} \Biggl[ 
\log \left[ {\sigma_5^{dS}+k_5\sin\zeta_0 \over \sigma_5^{dS}+k_5} \right] 
-{1 \over 2}  \left({k_5\cos\zeta_0 \over k_4^{dS}}\right)^2 +{k_5 \sigma_5^{dS} \over (k_4^{dS})^2}
\left( 1-\sin\zeta_0 \right) \Biggr]
\ee
Equation (\ref{matchingcond}) in five dimensions enables us to replace
the trigonometric functions  using:
\bml
\bea
\cos\zeta_0 &=& {2\lambda \over  1+\lambda^2} \\
\sin\zeta_0 &=& {1- \lambda^2  \over 1+\lambda^2}
\eea
\eml
where 
\be
\lambda={\mu_5\bar{\sigma}_5 \over 2k_4^{dS}}
\ee
This leads to a complicated expression. It is perhaps more
instructive to examine the behaviour for small $\mu$ \ie\ in the regime
where we have a vortex with a low energy density. In this regime we
find that:
\be \label{Bsmallmu}
B={256\pi^5\over (k^{dS}_4)^6}(G_5 \bar{\sigma}_5)^3\mu^4
+\mathcal{O} \rm (\mu^5)={256\pi^5\over (k^{dS}_4)^6}(\bar{G}_4)^3\mu^4
+\mathcal{O} \rm (\mu^5)
\ee
where $\bar{G}_4$ is the average of the four dimensional Newton's 
constants on the flat wall and the de Sitter wall. The presence of 
this average as opposed to a single four dimensional Newton's constant 
is due to the difference in brane tension on either side of the vortex. 
From equation (\ref{newton}) we see that this induces a difference in 
the Newton's constants on each wall. 

We now compare this to the result we would have got had we assumed no
extra dimensions. The analogous probability term, $B^{\prime}$, is
calculated in ~\cite{CDL}. When the energy
density of the bubble wall, $\mu$,  is small, we now find
that\footnote{In order to reproduce equation (\ref{CDresult}) using
separating a flat bubble spacetime and a de Sitter spacetime whose
radius of curvature is ${1 \over k_4}$. Then substitute into the
relevant equations and take $\mu$ to be small.}:
\be \label{CDresult}
B^{\prime}={256\pi^5\over (k_4)^6}(G_4)^3\mu^4 +\mathcal{O} \rm (\mu^5)
\ee
If we associate $G_4$ in equation (\ref{CDresult}) with $\bar{G}_4$ in equation (\ref{Bsmallmu}) we see that the approach of ~\cite{CDL}, where no extra dimensions are
present,  yields exactly the same result to the braneworld setup, at least for small $\mu$. 

Before we move on to discuss alternative instanton solutions we should
note that in the above analysis we have assumed ${\pi \over 2}>\zeta_0>0$. The
bounce solution presented is therefore really only valid if we have
$\lambda<1$. However, the extension to regions where $\lambda>1$
corresponds to allowing $\zeta_0$ to take negative values and
everything holds.

\subsection{Nucleation of an adS bubble spacetime in a flat false vacuum} 

\begin{figure}
\begin{center}
\includegraphics[width=8cm]{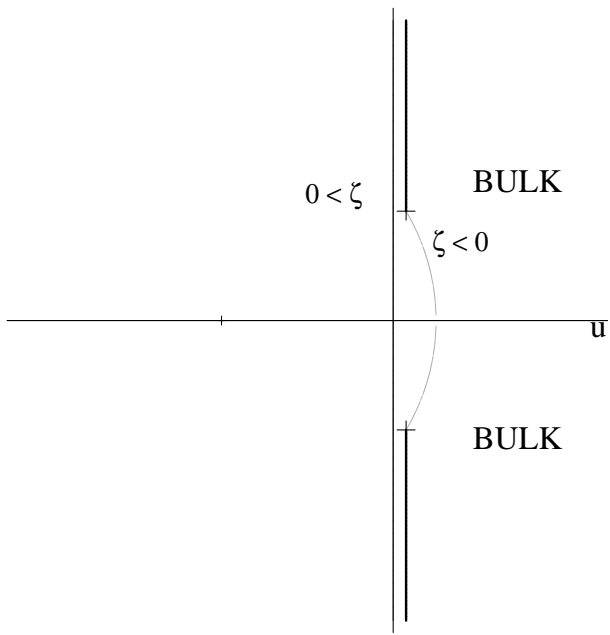}
\vskip 5mm
\caption{An example of a subcritical-critical wall ``bounce'' solution. 
This looks like an adS  bubble spacetime has nucleated on a flat wall.} 
\label{fig:bounce2}
\end{center}
\end{figure}

We now turn our attention to the decay of a flat false vacuum, and
the nucleation of an adS bubble spacetime. The braneworld analogue of 
 the flat false vacuum is given by a critical wall of tension
$\sigma_n^{flat}=k_n$ with no vortex ($u=$ const). This 
decays into a new ``bounce'' configuration
given by a subcritical wall (tension $\sigma_n^{adS}<k_n$) patched onto a
critical wall (tension $\sigma_n^{flat}=k_n$). Again, in order to avoid
generating an unphysical vortex, we must patch together trajectories
in the following way:
\be
Z=\cases{ \zeta + {1 \over k_{n-1}^{adS}}\sinh\zeta_0 & $\zeta>0$ \cr
{1 \over k_{n-1}^{adS}}\sinh(k_{n-1}^{adS}\zeta +\zeta_0) &
$\zeta<0$ }
\ee
where $(k_{n-1}^{adS})^2=k_n^2-(\sigma_n^{adS})^2$. The vortex tension
$\mu$ is related to the constant $\zeta_0$ in the following way:
\be \label{matchingcond2}
{\mu_n\bar{\sigma}_n \over 2k_{n-1}^{adS}}=\textrm{coth}\zeta_0-
\textrm{cosech}\zeta_0
\ee
By patching together $u=$ const.\ and figure  \ref{fig:walltraj}(a) we
again obtain a geometrical picture of our bounce solution (see figure
\ref{fig:bounce2}).

As before we now consider the probability term, $B$, given by the
difference between the euclidean actions of the bounce and the false
vacuum.  We shall not go into great detail here as the calculation is
very similar to that in the previous section. We should emphasize, 
however, that the bounce action will include as before 
an Einstein-Hilbert action  with negative cosmological constant in the
bulk, a Gibbons Hawking surface term on each wall, and tension
contributions from  each  wall and the vortex. Again we find that the delta function in the extrinsic curvature of the brane exactly cancels off the contribution from the vortex tension. The false vacuum action
omits the adS wall and vortex contributions, and contains no delta functions from extrinsic curvature. Recall that 
in each case we have two
copies of the bulk spacetime because of $Z_2$-symmetry across the
wall.

This time, we find that our probability term, $B$, is given by:
\be
B={4k_n^{2-n}\Omega_{n-2} \over 16\pi G_n} \left( I_n+\left({1 \over
n-1} \right) \left({k_n\sinh\zeta_0 \over
k_{n-1}^{adS}}\right)^{n-1} \right)
\ee
where the integral $I_n$ is now given by:
\be \label{newJn}
I_n=\int_{-u_0+u_1}^{u_c}  du \quad  \left ( u_0{[\rho(u)]^{n-3} 
\over u^{n-1}}+{[\rho(u)]^{n-1} \over u^n} \right )
\ee
where $u_c$ is an arbitrary constant corresponding to the position of
the flat wall, and
\bea
u_0 &=& {\sigma_n^{adS} \over (k_{n-1}^{adS})^2}(\sigma_n^{dS} +
k_n\cosh\zeta_0)u_c \\
u_1 &=& {k_n \over \sigma_n^{adS}}u_0 \\
\rho(u) &=& \sqrt{u_1^2-(u+u_0)^2} 
\eea
Again, although we could in principle solve this integral for any
positive integer $n$, we shall restrict our attention to $n=5$. In
this case we now find that:
\be
B  =  {8\pi^2k_5^{-3} \over 16\pi G_5}
\Biggl[ -\log\left[{\sigma_5^{adS}+k_5\cosh\zeta_0 \over 
\sigma_5^{adS}+k_5} \right] +{1 \over
2}  \left({k_5\sinh\zeta_0 \over
k_4^{adS}}\right)^2  + {k_5 \sigma_5^{adS} \over
(k_4^{adS})^2}\left(1-\cosh\zeta_0 \right) \Biggr]
\ee
We can now replace the hyperbolic functions using equation
(\ref{matchingcond2}):
\bml \bea
\sinh\zeta_0 &=& {2\lambda \over  1-\lambda^2} \\
\cosh\zeta_0 &=& {1+\lambda^2  \over 1-\lambda^2}
\eea \eml
where 
\be
\lambda={\mu_5\bar{\sigma}_5 \over 2k_4^{adS}}
\ee
This is again an ugly expression. It is more interesting to examine 
the behaviour at small $\mu$:
\be \label{newBsmallmu}
B={256\pi^5\over (k^{adS}_4)^6}(G_5 \bar{\sigma}_5)^3\mu^4
+\mathcal{O} \rm (\mu^5)={256\pi^5\over (k^{adS}_4)^6}(\bar{G}_4)^3\mu^4
+\mathcal{O} \rm (\mu^5)
\ee
This is  very similar to what we had for the nucleation of a flat
bubble spacetime in a de Sitter false vacuum with $\bar{G}_4$  now 
representing the average of the Newton's constants on the adS wall 
and the flat wall. Again we compare this to
the result from ~\cite{CDL} where we have no extra dimensions. When the
energy density of the bubble wall is small, the expression for the
probability term is again given by equation (\ref{CDresult}), where
${1 \over k_4}$ now corresponds to the radius of curvature of the 
adS spacetime. Once again we see that the braneworld result agrees exactly with ~\cite{CDL} in the small $\mu$ limit, provided we associate $G_4$ with $\bar{G}_4$.

Note that again we have assumed $\zeta_0>0$ and therefore, the
bounce solution given here is only valid for $\lambda<1$. The
extension to $\lambda>1$ is more complicated than for the nucleation
of the flat bubble in the previous section.  We now have to patch
together figure  \ref{fig:walltraj}(b) and figure \ref{fig:walltraj}(c). 
However, in ~\cite{CDL}, the
analogue of $\lambda>1$ violates conservation of energy as one tunnels
from the false vacuum  to the new configuration. In the braneworld set up we
should examine what happens as $\lambda$ approaches unity from
below. In this limit, $\zeta_0$ becomes infinite, and the adS bubble
encompasses the entire brane. The probability,  $\mathcal{P}$, of
this happening is zero and so there is no vacuum decay. Beyond this, in
analogy with ~\cite{CDL}, we would suspect that the energy of the 
false vacuum is
insufficient to allow the nucleation of a bubble with a large wall
energy density.  This is indeed the case. When we calculate the
probability term, $B$,  for the adS bubble in a flat, spherical 
false vacuum, we find that it is divergent and the probability of bubble 
nucleation vanishes. This divergence comes from the fact that the 
false vacuum touches the adS boundary whereas the bounce solution does not.
 
Finally, we could also have created an adS bubble in a flat spacetime using a
$\kappa=0$ vortex.  However, it is of no interest since the probability of
bubble nucleation is exponentially
suppressed by the vortex volume.

\subsection{Ekpyrotic Instantons}

The notion of the Ekpyrotic universe \cite{EKP} proposes that 
the Hot Big Bang came about as the result of a collision between 
two braneworlds. The model claims to solve many of the problems facing 
cosmology without the need for inflation. Although the authors work 
mainly in the context of heterotic M theory, they acknowledge that an 
intuitive understanding can be gained by considering Randall-Sundrum type 
braneworlds. In this context, we regard the pre-Big Bang era in the 
following way. We start off with two branes of equal and opposite tension: 
the hidden brane of positive tension, $\sigma$, and the visible brane of 
negative tension, $-\sigma$.  A bulk brane with a small positive 
tension, ${\epsilon \over 2}$, then ``peels off'' the hidden brane 
causing its tension to fall to $\sigma-\epsilon$. The bulk brane is then 
drawn towards our universe, the visible brane, until it collides with us, 
giving rise to the Hot Big Bang. 

\begin{figure} 
\centerline{\epsfig{file=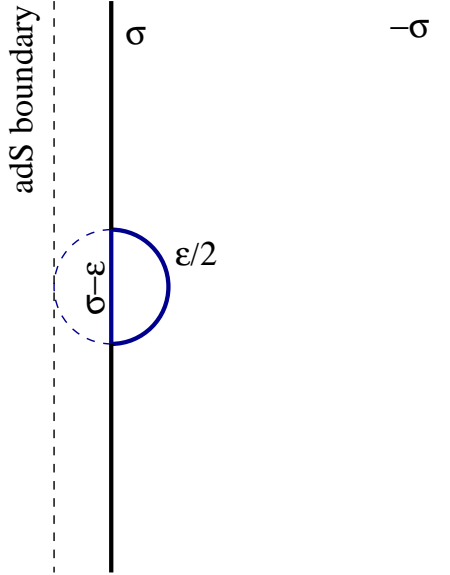,width=6cm}}
\vskip 5mm
\caption{ The Ekpyrotic Instanton}
\label{fig:Ekpyrotic}
\end{figure}

The process of ``peeling off'' is not really considered in great detail 
in ~\cite{EKP}. They suggest that the hidden brane undergoes a 
small instanton transition with the nucleation of a bubble containing 
a new hidden brane with decreased tension, and the bulk brane. 
The walls of this bubble then expand at the speed of light until it 
envelopes all of the old hidden brane. Given that all the branes in 
this model are critical we can illustrate the instanton solution in the
simplified RS set-up by using 
a suitable combination of critical wall solutions. In conformal 
coordinates, critical walls look like planes ($u=\text{const}$) or 
spheres (see figure \ref{fig:walltraj}(c)). In describing the Ekpyrotic 
instanton we present the visible and hidden branes (old and new) as planes. 
The bulk brane is given by a sphere that intersects the hidden brane, 
separating the old and new branches (see figure \ref{fig:Ekpyrotic}).

Given this geometrical picture we can calculate the probability 
of tunneling to this configuration from the initial two brane state. 
We proceed much as we did in the previous section, and obtain the 
following expression for the probability term:
\be
B={\pi^2 \epsilon \over 4}\left({3 \over k^3}\ln(1+k^2Z_0^2)
-{2k^{-2}Z_0^2 \over 1+k^2Z_0^2}-{Z_0^2 \over k^2}-{Z_0^4 \over4} \right) 
+\mathcal{O} \rm (\epsilon^2)
\ee
where $k$ is related to the cosmological constant in the bulk of the 
initial state ($\Lambda = -6k^2$), and $Z_0$ is a free parameter 
related to the ``size'' of the bubble: the larger the value of $Z_0$, 
the larger the bubble.  We should not be worried by this freedom 
in $Z_0$, as we are working with Randall-Sundrum  braneworlds which are 
much simpler than the M5 branes of heterotic M theory. When we return 
to the  M theory context, we lose a number of degrees of freedom and 
one might expect the value of $Z_0$ to be fixed. However, since we 
are dealing with a ``small' instanton, we might expect $Z_0$ itself 
to be small, and the probability term approximates to the following:
\be
B={\pi^2 \over 16}\epsilon Z_0^4 + \mathcal{O} \rm (\epsilon^2, Z_0^6)
\ee
We should once again stress however, that this is an extremely simple
and naive calculation that ignores any dynamics of the additional scalars,
or other fields, that result from a five-dimensional heterotic M-theory 
compactification \cite{HM5}.
Another point to note is that while we can have a small brane peel off from the
positive tension braneworld, we cannot have one peel off from the negative
tension braneworld, as a quick glance at figure \ref{fig:Ekpyrotic} shows.
Such a brane, being critical, must have the form of a sphere grazing 
the adS boundary, which therefore necessarily would intersect the positive
tension brane as well. 

\section{Discussion}\label{sect:disc}

In this paper we have derived the general spacetime of an infinitesimally
thin braneworld containing a nested domain wall or domain ribbon.
These results can be regarded as a ``zero-thickness'' limit of some
thick braneworld (and ribbon) field theory model, in much the same way 
as the Randall-Sundrum model can be regarded as the limit of a thick brane. 
The ribbon can therefore be interpreted as a kink in a scalar condensate 
living on the brane.  Cosmologically, this scalar plays the role of an 
inflaton, which can therefore potentially give rise to an old inflationary 
model.  In section \ref{sect:inst} we determined the instanton for false vacuum
decay, however, we found that the presence of the extra dimension does not 
ameliorate the usual problems afflicting old inflation. 

Given that these results were obtained in the absence of the adS soliton c-term, a natural question might be to ask how they are modified for $c \neq 0$, in the canonical metric (\ref{canbulk}). The adS soliton was originally investigated by Horowitz and Myers \cite{HM}\ in the context of a modified positive
energy conjecture in asymptotically locally adS spacetimes, and 
has also been used to obtain braneworld solutions in six
dimensions \cite{CLP,LMW}. These models slice the flat ($\kappa=0$)
soliton cigar in six dimensions at fixed `radius' -- which
corresponds to a source consisting of a superposition of a 4-brane and a
smeared 3-brane (\ie\ smearing out our $\delta$-function source throughout
the 4-brane). In addition these papers also included 3-branes
located at the euclidean horizon $Z_+$, via conical deficits in the 
$R$-coordinate.

Clearly we can use our formalism to re-localize the 3-brane of
\cite{CLP,LMW}, as well as derive more general braneworlds with
arbitrary $\kappa$, tension $\sigma$, and overall spacetime dimension $n$.
Leaving a detailed study for later work, we now simply remark on
the qualitative behaviour of such braneworlds. The
equations of motion (\ref{srctraject}) for a $Z_2$-symmetric
trajectory become:
\bml\label{adssoleom}\bea
Z^{\prime 2} &=& aZ^2 + \kappa - {c\over Z^{n-3}}\label{adssola}\\
Z'' &=& a Z + \left({n-3 \over 2} \right){c\over Z^{n-2}} - 4\pi G_n \sum_i
\mu_i\sigma_n Z\delta(\zeta - \zeta_i)\label{adssolb}\\
R' &=& {\sigma_n Z^{n-2}\over k_n^2 Z^{n-1} + \kappa Z^{n-3} - c} 
\label{adssolc}
\eea\eml
where $a = k_n^2 - \sigma_n^2$ as before, and (\ref{adssolb}) now allows for
a multitude of domain ribbons of tension $\mu_i$ located at $\zeta_i$.

We can quickly see the qualitative behaviour of solutions to these equations.
The generic trajectory (which must be periodic in $R$) will consist of 
two segments of $Z(\zeta)$ of opposite gradient which patch together at
a positive tension domain ribbon at $R=0$, say, and a 
negative tension domain ribbon at $R=\Delta R/2$. This is exactly analogous
to the usual situation with a domain wall spacetime when we need both
positive and negative tension walls to form a compact extra dimension. 
However, we see that with the bulk ``mass'' term, there are now also other
possibilities, since (\ref{adssola}) now has at least one root for
$Z>Z_+$, and in fact two for supercritical walls. This root corresponds
to a zero of $Z'$, and hence the possibility for a smooth transition from
the positive branch of $Z'$ to the negative branch. We can therefore form 
a trajectory which loops symmetrically around the cigar, and has only
one kink -- which we can fix to be a positive tension vortex. Of course,
the tension of this vortex will be determined by the other parameters
of the set-up: the bulk mass, cosmological constant and the braneworld
tension, but this is no worse a fine tuning than is already
present in conventional RS models. Note that this is now distinct from
a domain wall on a compact extra dimension, as we can construct a
domain ribbon spacetime with only a single positive tension ribbon on
all braneworld geometries \ie\ with positive (supercritical), zero
(critical) or negative (subcritical) induced cosmological constants.
In addition, for a supercritical braneworld (\ie\ a positive induced
cosmological constant) we have the possibility of dispensing
with the ribbon altogether, and having a smooth trajectory with two
roots of $Z'$, where the braneworld smoothly wraps the cigar, although
this requires a fine tuned mass term.

In all cases the induced braneworld geometry has the form
\be
ds^2 = Z^2(\zeta) \left [ dX_{n-2}^2 \right] _\kappa - d\zeta^2
\ee
(where of course $\zeta$ has a finite range). The energy momentum
tensor of this spacetime is
\bml\bea
8\pi G_{n-1} T^\mu_\nu &=& \left[ -(n-3){Z''\over Z} - {(n-3)(n-4)\over2}
{(Z^{\prime2} - \kappa)\over Z^2} + {(n-3)\over2} \mu_n\sigma_n\delta(\zeta) \right]\delta^{\mu}_{\nu}
\nonumber\\
&=&  - {(n-3)\over2} \left [ (n-2)a + {c\over Z^{n-1}} -
\mu_n\sigma_n\delta(\zeta) \right ]\delta^{\mu}_{\nu}\\
8\pi G_{n-1} T^\zeta_\zeta &=& - {(n-2)(n-3)\over2}
{(Z^{\prime2} - \kappa)\over Z^2} = - {(n-2)(n-3)\over2}
\left ( a - {c\over Z^{n-1}} \right)
\eea\eml
which has three distinct components: A cosmological constant (the $a$-term)
which reflects the lack of criticality of the braneworld when it is
nonvanishing. The domain ribbon terms ($\mu_i$) when present indicate
the presence of a nested $(n-3)$-brane -- note the normalization is
precisely correct for the induced $(n-1)$-dimensional Newton's constant.
And the final $c$-term is in fact a negative stress-energy tensor, which
is of Casimir form, and can be directly associated to the Casimir 
energy of the CFT living in the braneworld, in the same way as the
stress-tensor of the radiation cosmology obtained by slicing a standard
Schwarzschild-adS black hole can be interpreted as the energy of the
CFT at finite Hawking temperature on the braneworld \cite{GVS}.

\section*{Acknowledgements}

We would like to thank Ian Davies, James Gregory, Clifford Johnson,
Ken Lovis, and David Page for useful discussions. 
R.G.\ was supported by the Royal Society and A.P.\ by PPARC.

\appendix

\section{Details of Instanton Calculation}

In section IV we calculated the probability of bubble nucleation 
in a number of braneworld situations. The details of these calculations 
are remarkably similar for both the flat bubble and the AdS bubble. 
In this Appendix we shall present the calculation for the flat bubble 
spacetime forming in a de Sitter false vacuum.

Consider now equations (\ref{bounceaction}a-d). Our solution 
satisfies the equations of motion both in the bulk and on the brane. 
The bulk equations of motion are just Einstein's equations (in 
euclidean signature) with a negative cosmological constant:
\be \label{einstein}
R_{ab}-{1\over 2}Rg_{ab}=-\Lambda g_{ab}
\ee
from which we can quickly obtain 
\be\label{bulkint}
R-2\Lambda={4\Lambda \over n-2}=-2(n-1)k_n^2
\ee
where we have used the relation (\ref{lamkn}). 
The brane equations of motion are just the Israel equations given that 
we have a brane tension and a nested domain wall:
\be \label{Israel}
\Delta K_{\mu\nu}-\Delta K h_{\mu\nu}=8\pi G_n\sigma h_{\mu\nu} 
+8\pi G_n\mu \delta(\zeta) \gamma_{\mu\nu}
\ee
where $\sigma$ is $\sigma^{flat}$ and $\sigma^{dS}$ on the flat 
and de Sitter walls respectively. We can therefore read off 
the following expression:
\be\label{Kdiff}
\Delta K=-2(n-1)\sigma_n-\mu_n \delta(\zeta)
\ee
where we have also used (\ref{rescalesigmu}). 
We are now ready to calculate the action. Inserting (\ref{bulkint}) and 
(\ref{Kdiff}) in (\ref{bounceaction}), 
we immediately see that the contribution from the vortex is cancelled 
off by the delta function in the extrinsic curvature and we are left with
\be
S_{bounce}={2(n-1)k_n^2\over 16\pi G_n}\int_{bulk} d^n x \sqrt{g}
-{4\sigma_n^{flat}\over 16\pi G_n}\int_{flat} d^{n-1} x \sqrt{h}
-{4\sigma_n^{dS}\over 16\pi G_n}\int_{dS} d^{n-1} x \sqrt{h}
\ee

The expression for $S_{false}$ is similar except that there is of course no 
flat wall contribution and the limits for the bulk and de Sitter brane 
integrals run over the whole of the de Sitter sphere interior and surface
respectively.

Working in euclidean conformal coordinates (\ie\ the metric (\ref{confmet})
rotated to euclidean signature) the bulk measure is simply
\be \label{bulkmeasure}
\sqrt{g} d^n x= {\rho^{n-2} \over (k_nu)^n}
du\, d\rho\, d \Omega_{n-2}
\ee
where $d\Omega_{n-2}$ is the measure on a unit $n-2$ sphere.
From (\ref{hyper}) and (\ref{dswallrelations}), the de Sitter wall 
is given by 
\be \label{dSwall}
(u-u_0)^2+\rho^2= u_1^2
\ee
so the induced metric on this wall is given by:
\be \label{dSinduced}
ds^2_{n-1}={1 \over k_n^2u^2} \left [ \left ( 
{k_n u_0 \over \sigma^{dS}_n \rho(u)} \right )^2 du^2
+\rho(u)^2d\Omega_{n-2} \right]
\ee
where $\rho(u)$ is given in (\ref{dswallrelc}). 
As we did for the bulk, we can now read off the de Sitter wall measure:
\be \label{dSmeaure}
\sqrt{h} d^{n-1} x=  u_1{ \rho(u)^{n-3}\over (k_nu)^{n-1}}\, du\, 
d\Omega_{n-2}\,.
\ee
Now consider the flat wall. This is given by $u=u_c$ where $u_c$ is 
given by (\ref{dswallrela}), and the measure can be easily seen
to be
\be \label{flatmeasure}
\sqrt{h} d^{n-1}x = { \rho^{n-2}\over (k_nu_c)^{n-1}}\, d\rho\,d\Omega_{n-2}.
\ee

Now we are ready to evaluate the probability term $B=S_{bounce}-S_{false}$. 
Given each of the measures we have just calculated and taking care 
to get the limits of integration right for both the bounce action 
and the false vacuum action, we arrive at the following expression:
\bea \label{bounceexpression}
B &=& -{4(n-1)k_n^2\over 16\pi G_n}\Omega_{n-2}\int_{u_0-u_1}^{u_c}du  
\int_0^{\rho(u)} d \rho  {\rho^{n-2} \over (k_nu)^n} \nonumber \\
& & -{4\sigma_n^{flat}\over 16\pi G_n}\Omega_{n-2}\int_0^{\rho(u_c)}
d \rho { \rho^{n-2}\over (k_nu_c)^{n-1}}
+{4\sigma_n^{dS}\over 16\pi G_n}\Omega_{n-2}\int_{u_0-u_1}^{u_c} d u  
u_1{ \rho(u)^{n-3}\over (k_nu)^{n-1}} \nonumber \\
\eea
We should note that we have a factor of two in the bulk part of the 
above equation arising from the fact that we have two copies of the 
bulk spacetime. If we use the fact that:
\be
\rho(u_c)={k_n u_c \over k_{n-1}^{dS}}\cos\zeta_0
\ee
along with $\sigma_n^{flat}=k_n$ and equation (\ref{dswallrelb}), we 
can simplify (\ref{bounceexpression}) to arrive at 
equation (\ref{probterm}).

\end{document}